\documentclass[prc,twocolumn,amsmath,amssymb,floatfix,nofootinbib]{revtex4}
\usepackage{mathrsfs,bm}
\usepackage{longtable,lscape}
\usepackage{txfonts}
\usepackage{indentfirst}
\usepackage{graphicx,color,dcolumn,booktabs}
\usepackage{multirow}
\usepackage{indentfirst}
\usepackage{multirow}
\usepackage{epsfig}
\usepackage{graphicx,color,dcolumn}
\usepackage{epstopdf}
\usepackage{amsmath}
\usepackage{multirow}
\usepackage{diagbox}
\usepackage{changes}
\usepackage{threeparttable}

\usepackage{color}
\usepackage{amssymb}
\definecolor{cover}{rgb}{0.77,0.87,0.88}
\definecolor{blueone}{rgb}{0.1,0.1,.7}
\definecolor{citec}{rgb}{0.14,0.47,0.09}
\definecolor{two}{rgb}{0.0,0.5,0.}
\definecolor{three}{rgb}{.5,.1,0.15}
\usepackage[bookmarks=true,bookmarksopen=false,plainpages=false,breaklinks=true,
   bookmarksnumbered=true,hypertexnames=false,
   filecolor=blue,urlcolor=three,menucolor=three,
   linkcolor=three,citecolor=blueone, colorlinks,
   anchorcolor=blue,runcolor=pink,frenchlinks=red
   pdfstartview=FitH,pdftitle=title,%
   pdfauthor=author]{hyperref}

\usepackage{relsize}
\usepackage{xspace}
\def\babar{\mbox{\slshape B\kern-0.1em{\smaller A}\kern-0.1em
    B\kern-0.1em{\smaller A\kern-0.2em R}}}

\allowdisplaybreaks[4]
\usepackage{color}
\begin{document}
\title{$P_{cs}(4459)$ and other possible molecular states from $\Xi_{c}^{(*)}\bar{D}^{(*)}$ and $\Xi'_c\bar{D}^{(*)}$  interactions}

\author{Jun-Tao Zhu, Lin-Qing Song, Jun He\footnote{Corresponding author: junhe@njnu.edu.cn}}
\affiliation{$^1$Department of  Physics and Institute of Theoretical Physics, Nanjing Normal University, Nanjing 210097, China}

\date{\today}
\begin{abstract}

Recently, the LHCb Collaboration reported a new structure $P_{cs}(4459)$ with a mass of 19~MeV below the $\Xi_c \bar{D}^{*} $ threshold. It may be a candidate of molecular state from the $\Xi_c \bar{D}^{*} $ interaction. In the current work, we perform a  coupled-channel  study of the $\Xi_c^*\bar{D}^*$, $\Xi'_c\bar{D}^*$, $\Xi^*_c\bar{D}$, $\Xi_c\bar{D}^*$, $\Xi'_c\bar{D}$, and $\Xi_c\bar{D}$  interactions in the quasipotential Bethe-Salpeter equation approach. With the help of  the heavy quark chiral effective Lagrangian, the potential is constructed by light meson exchanges. Two $\Xi_c \bar{D}^{*} $ molecular states are produced with spin parities $ J^P=1/2^-$ and  $3/2^- $.  The lower state with $3/2^-$  can be related to the observed $P_{cs}(4450)$ while two-peak structure cannot be excluded.  Within the same model, other strange hidden-charm pentaquarks are also predicted.  Two states with spin parities $1/2^-$ and $3/2^-$ are predicted near the $\Xi'_c\bar{D}$, $\Xi_c\bar{D}$,  and  $\Xi_c^*\bar{D}$ thresholds, respectively.  As two states near $\Xi_c \bar{D}^{*}$ threshold, two states are produced with $1/2^-$ and $3/2^-$  near the $\Xi'_c\bar{D}^*$ threshold. The couplings of the molecular states to the considered channels are also discussed.  The experimental  research of those states are helpful to understand the origin and internal structure of the $P_{cs}$ and $P_c$ states.

\end{abstract}

\maketitle
\section{INTRODUCTION}

Recently, the LHCb Collaboration released their results about the $\Xi_b \rightarrow J/\psi K^-\Lambda $ decay, which indicates a new resonance structure named $P_{cs}(4459)$, which carries a mass of $4458.8 \pm 2.9^{+4.7}_{-1.1}$~MeV and a width of $17.3\pm 6.5 ^{+8.0}_{-5.7}$~MeV~\cite{Aaij:2020gdg}.
Such structure  is just $19$~MeV below the $\Xi_c \bar{D}^{*} $ threshold and the width is relatively narrow.
It follows a series of observations about the hidden-charm pentaquarks. In 2015, the LHCb Collaboration firstly reported two structures $P_c(4380)$ and $P_c(4450)$ in the $\Lambda_b\to J/\psi K^-p$~\cite{Aaij:2015tga}, which are close to the $\Sigma^*_c\bar{D}$ and $\Sigma_c\bar{D}^*$ thresholds, respectively.  Such observation confirmed previous predictions about the hidden-charm pentaquarks~\cite{Wu:2010jy,Yang:2011wz,Wang:2011rga,Xiao:2013yca}.  Due to the closeness to the thresholds, these structures were soon interpreted as molecular states in the literature~\cite{Chen:2015loa, Chen:2015moa,Karliner:2015ina,Roca:2015dva,He:2015cea,Burns:2015dwa}. Other pictures, such as compact pentaquark ~\cite{Yuan:2012wz, Lebed:2015tna,Maiani:2015vwa} and anomalous triangle singularity~\cite{Liu:2015fea}, were also proposed to explain the observation.
In 2019, the LHCb Collaboration updated their results.  The upper $P_c(4450)$ was found to be a two-peak structure named as $P_c(4457)$ and $P_c(4440)$, and a new structure, $P_c(4312)$, was observed near $\Sigma_c\bar{D}$ threshold~\cite{Aaij:2019vzc}. The new result strongly supports the molecular state picture. If  we include the $P_c(4380)$ state, which was not confirmed but also not  excluded by new observation, the four states provide a wonderful spectrum of  S-wave states from the $\Sigma_c\bar{D}-\Sigma_c^*\bar{D}-\Sigma_c\bar{D}^*$ interaction.

The observation of the hidden-charm pentaquarks inspires a large amount of  theoretical studies about their internal structure~\cite{Chen:2019asm,Chen:2019bip,Fernandez-Ramirez:2019koa,Wu:2019rog,Huang:2013mua,Wang:2019got,Huang:2019jlf,Liu:2019tjn,He:2019ify,He:2019rva, Lin:2019qiv,Xiao:2019mst}. The partners of the hidden-charm pentaquarks were also widely discussed  in the literature~\cite{He:2015yva, He:2017aps,Lin:2018kcc,Wu:2017weo,Wang:2019zaw,Huang:2018wed,Yang:2018oqd,Wang:2019ato,Zhu:2020vto}.   Replacing one of the light quark by strange quark, one can naturally obtain a strange hidden-charm pentaquark, which have been studied in the pioneer work~\cite{Wu:2010jy} and some studies after the observation of $P_{c}$ states~\cite{Anisovich:2015zqa,Wang:2015wsa,Feijoo:2015kts,Chen:2015sxa,Chen:2016ryt,Lu:2016roh,Xiao:2019gjd,Zhang:2020cdi,Wang:2019nvm}. The observation of these partners is very helpful to understand the internal structure of the pentaquarks.

The recent observation  of  $P_{cs}(4459)$ favors the molecular state interpretations of the hidden-charm pentaquarks~\cite{Chen:2020uif,Peng:2020hql,Wang:2020eep,Chen:2020kco}. It can be taken as the strange partner of the $P_c(4450)$.  According  to its mass, it can be assigned as the candidate of  a $\Xi_c \bar{D}^{*}$ state with spin parity $J^P=1/2^-$ or $3/2^-$.  Due to lack of the partial-wave treatment, the current experimental analysis did not provide the information about its spin parity. From the experience of study about the $P_c(4450)$, one can expect that there may be two structures in the $P_{cs}(4459)$. In Ref.~\cite{Wang:2019nvm}, the author predicted two $\Xi_c \bar{D}^{*}$ states with  $J^P=1/2^-$ and $3/2^-$ which carry masses of  $4456.9^{+3.2}_{-3.3}$ and $4463.0^{+2.8}_{-3.0}$~MeV, respectively. Different mass order of two $\Xi_c \bar{D}^{*}$ states was also suggested in the literature. In Ref.~\cite{Peng:2020hql}, the calculation yields masses of $4469$ and $4453-4463$ MeV for states with ${1}/{2}$ and ${3}/{2}$, respectively.

In our previous work, the hidden-charm pentaquarks were well explained as molecular states in the quasipotential Bethe-Salpeter equation (qBSE) approach~\cite{He:2019ify,He:2019rva}, which was extended  to predict the hidden-bottom pentaquarks~\cite{Zhu:2020vto}. Such studies provide a good theoretical frame to study the strange  hidden-charm pentaquark. Due to the heavy quark symmetry, the model has been constrained by experimentally observed hidden-charm pentaquarks. Hence, it is interesting to perform a calculation about the molecular states from the $\Xi_{c}^{(*,')}\bar{D}^{(*)}$ interaction in the same frame. The strange hidden-charm pentaquarks are also possibly produced from the interactions of  charmed baryon and strange anticharm meson. Compared with the couplings between channels in the  $\Xi_{c}^{(*,')}\bar{D}^{(*)}$ interactions where the pion exchange is possible, the couplings between two types of interactions though kaon exchange should be weaker.  Hence, in the current work, we consider six channels in $\Xi_{c}^{(*,')}\bar{D}^{(*)}$ interaction, which are related to the $P_{cs}(4459)$.

With the help of the heavy quark chiral effective Lagrangians, which was also adopted to reproduce the hidden-charm pentaquarks,  the one-boson-exchange model will be adopted in the current work to construct the interaction kernel to calculate the scattering amplitude by solving the qBSE.  The molecular states can be studied by searching for the  poles of  the complex energy plane. The coupled-channel effects will be also included explicitly to produce the widths of the molecular states. The molecular state interpretation of the $P_{cs}(4459)$ will be discussed based on the results. More molecular states will be  also predicted, which can be searched in the future experiment.

This article is organized as follows. After introduction, Section~\ref{Sec: Formalism} shows the details of dynamics of the $\Xi_c^*\bar{D}^*$, $\Xi'_c\bar{D}^*$, $\Xi^*_c\bar{D}$, $\Xi_c\bar{D}^*$, $\Xi'_c\bar{D}$, and $\Xi_c\bar{D}$ interactions, including the relevant effective Lagrangians, reduction of potential kernel and a brief introduction of the qBSE. In Section~\ref{3}, the results with single-channel calculation are given first. Then, coupled-channel results are presented, and the importance of the channels considered  are discussed. Finally, summary  and  discussion are given in Section~\ref{4}. %Finally, summary  and  discussion will be given in section~\ref{5}.

\section{Theoretical frame}\label{Sec: Formalism}

In the current work, we follows the same theoretical frame in the study of the LHCb hidden-charm pentaquarks~\cite{He:2019ify,He:2019rva},   to study the strange hidden-charm pentaquark. The one-boson-exchange potential of the interactions of strange charmed baryon $\Xi_c^{(*,')}$ and anticharm meson $\bar{D}^{(*)}$ is constructed as  dynamical kernel. The pseudoscalar $\mathbb{P}$, vector $\mathbb{V}$ and scalar $\sigma$  exchanges  will be considered, and the effective Lagrangian depicting the couplings of light mesons and anticharmed mesons $\bar{D}^{(*)}$ or strange charmed baryons $\Xi_c^{(*,')}$ are required and will be presented in the below.

\subsection{Relevant Lagrangians}
First, we consider the couplings of light mesons and heavy-light anticharmed mesons $\mathcal{P}=(\bar{D}^0, D^-, D^-_s)$.  Considering heavy quark limit and chiral symmetry,  the Lagrangians have been  constructed  in the literature as \cite{Cheng:1992xi,Yan:1992gz,Wise:1992hn,Casalbuoni:1996pg},
\begin{align}
\mathcal{L}_{HH\mathbb{P}}&= ig_{1}\langle \bar{H}_a^{\bar{Q}} \gamma_\mu
{\cal A}_{ba}^\mu\gamma_5 {H}_b^{\bar{Q}}\rangle,\nonumber\\
\mathcal{L}_{HH\mathbb{V}}&= -i\beta\langle \bar{H}_a^{\bar{Q}}  v_\mu
(\mathcal{V}^\mu_{ab}-\rho^\mu_{ab}){H}_b^{\bar{Q}}\rangle
+i\lambda\langle \bar{H}_b^{\bar{Q}}
\sigma_{\mu\nu}F^{\mu\nu}(\rho)\bar{H}_a^{\bar{Q}}\rangle,
\nonumber\\
\mathcal{L}_{ HH\sigma}&=g_s \langle \bar{H}_a^{\bar{Q}}\sigma
\bar{H}_a^{\bar{Q}}\rangle,\label{eq:lag1}
\end{align}
where  ${\cal A}^\mu=\frac{1}{2}(\xi^\dag\partial_\mu\xi-\xi \partial_\mu
\xi^\dag)=\frac{i}{f_\pi}\partial_\mu{\mathbb P}+\cdots$ is the axial current with
$\xi=\exp(i\mathbb{P}/f_\pi)$ and $f_\pi=132$
MeV.
The vector current $\mathcal{V}_\mu=\frac{i}{2}[\xi^\dag(\partial_\mu\xi)
+(\partial^\mu\xi)\xi^\dag]=0$, and vanishes.
$\rho^\mu_{ba}=ig_\mathbb{V}\mathbb{V}^\mu_{ba}/\sqrt{2}$, and
$F^{\mu\nu}(\rho)=\partial_\mu \rho_\nu - \partial_\nu \rho_\mu +
[\rho_\mu,{\ } \rho_\nu]$. The $\mathbb
P$ and $\mathbb V$ are the pseudoscalar and vector matrices as
\begin{align}
    {\mathbb P}&=\left(\begin{array}{ccc}
        \frac{1}{\sqrt{2}}\pi^0+\frac{\eta}{\sqrt{6}}&\pi^+&K^+\\
        \pi^-&-\frac{1}{\sqrt{2}}\pi^0+\frac{\eta}{\sqrt{6}}&K^0\\
        K^-&\bar{K}^0&-\frac{2\eta}{\sqrt{6}}
\end{array}\right),\nonumber\\
\mathbb{V}&=\left(\begin{array}{ccc}
\frac{\rho^{0}}{\sqrt{2}}+\frac{\omega}{\sqrt{2}}&\rho^{+}&K^{*+}\\
\rho^{-}&-\frac{\rho^{0}}{\sqrt{2}}+\frac{\omega}{\sqrt{2}}&K^{*0}\\
K^{*-}&\bar{K}^{*0}&\phi
\end{array}\right).\label{MPV}
\end{align}
The doublet are defined as $H_a^{\bar{Q}}=[\mathcal{P}^{*\mu}_a\gamma_\mu-\mathcal{P}_a\gamma_5]\frac{1-\rlap\slash
v}{2}$ and
$\bar{H}=\gamma_0H^\dag\gamma_0$.
The
$\mathcal{P}
$ and $\mathcal{P}^*%(\widetilde{\mathcal{P}}^*)
$ satisfy the normalization relations $\langle
0|{\mathcal{P}}|\bar{Q}{q}(0^-)\rangle
%=\langle 0|\widetilde{\mathcal{P}}|\bar{Q}q(0^-)\rangle
=\sqrt{M_\mathcal{P}}$ and $\langle
0|{\mathcal{P}}^*_\mu|\bar{Q}{q}(1^-)\rangle=%\langle
%0|\widetilde{\mathcal{P}}^{*}_\mu|\bar{Q}q(1^-)\rangle
\epsilon_\mu\sqrt{M_{\mathcal{P}^*}}$.

The Lagrangians can be further expanded
as follows for explicitly application,
\begin{align}
  \mathcal{L}_{\mathcal{P}^*\mathcal{P}\mathbb{P}} &=
 i\frac{2g\sqrt{m_{\mathcal{P}} m_{\mathcal{P}^*}}}{f_\pi}
  (-\mathcal{P}^{*\dag}_{a\lambda}\mathcal{P}_b
  +\mathcal{P}^\dag_{a}\mathcal{P}^*_{b\lambda})
  \partial^\lambda\mathbb{P}_{ab},\nonumber\\
    \mathcal{L}_{\mathcal{P}^*\mathcal{P}^*\mathbb{P}} &=
-\frac{g}{f_\pi} \epsilon_{\alpha\mu\nu\lambda}\mathcal{P}^{*\mu\dag}_a
\overleftrightarrow{\partial}^\alpha \mathcal{P}^{*\lambda}_{b}\partial^\nu\mathbb{P}_{ba},\nonumber\\
    \mathcal{L}_{\mathcal{P}^*\mathcal{P}\mathbb{V}} &=
\sqrt{2}\lambda g_V\varepsilon_{\lambda\alpha\beta\mu}
  (-\mathcal{P}^{*\mu\dag}_a\overleftrightarrow{\partial}^\lambda
  \mathcal{P}_b  +\mathcal{P}^\dag_a\overleftrightarrow{\partial}^\lambda
 \mathcal{P}_b^{*\mu})(\partial^{\alpha}\mathbb{V}^\beta)_{ab},\nonumber\\
	\mathcal{L}_{\mathcal{P}\mathcal{P}\mathbb{V}} &= -i\frac{\beta	g_V}{\sqrt{2}}\mathcal{P}_a^\dag
	\overleftrightarrow{\partial}_\mu \mathcal{P}_b\mathbb{V}^\mu_{ab}, \nonumber\\
  \mathcal{L}_{\mathcal{P}^*\mathcal{P}^*\mathbb{V}} &= - i\frac{\beta
  g_V}{\sqrt{2}}\mathcal{P}_a^{*\dag}\overleftrightarrow{\partial}_\mu
  \mathcal{P}^*_b\mathbb{V}^\mu_{ab}\nonumber\\
  &-i2\sqrt{2}\lambda  g_Vm_{\mathcal{P}^*}\mathcal{P}^{*\mu\dag}_a\mathcal{P}^{*\nu}_b(\partial_\mu\mathbb{V}_\nu-\partial_\nu\mathbb{V}_\mu)_{ab}
,\nonumber\\
  \mathcal{L}_{\mathcal{P}\mathcal{P}\sigma} &=
  -2g_s m_{\mathcal{P}}\mathcal{P}_a^\dag \mathcal{P}_a\sigma, \nonumber\\
  \mathcal{L}_{\mathcal{P}^*\mathcal{P}^*\sigma} &=
  2g_s m_{\mathcal{P}^*}\mathcal{P}_a^{*\dag}
  \mathcal{P}^*_a\sigma,\label{LD}
\end{align}
where  the $v$ is replaced by $i\overleftrightarrow{\partial}/2\sqrt{m_im_f}$ with the $m_{i,f}$ is for the initial or final $\bar{D}^{(*)}$ meson.

The Lagrangians for the couplings between  charmed baryon and light mesons can also be constructed under the heavy quark limit and  chiral symmetry as,
\begin{align}
{\cal L}_{S}&=-
\frac{3}{2}g_1(v_\kappa)\epsilon^{\mu\nu\lambda\kappa}{\rm tr}[\bar{S}_\mu
{\cal A}_\nu S_\lambda]+i\beta_S{\rm tr}[\bar{S}_\mu v_\alpha (\mathcal{V}^\alpha-
\rho^\alpha)
S^\mu]\nonumber\\
& + \lambda_S{\rm tr}[\bar{S}_\mu F^{\mu\nu}S_\nu]
+\ell_S{\rm tr}[\bar{S}_\mu \sigma S^\mu],\nonumber\\
{\cal L}_{B_{\bar{3}}}&= i\beta_B{\rm tr}[\bar{B}_{\bar{3}}v_\mu(\mathcal{V}^\mu-\rho^\mu)
B_{\bar{3}}]
+\ell_B{\rm tr}[\bar{B}_{\bar{3}}{\sigma} B_{\bar{3}}], \nonumber\\
{\cal L}_{int}&=ig_4 {\rm tr}[\bar{S}^\mu {\cal A}_\mu B_{\bar{3}}]+i\lambda_I \epsilon^{\mu\nu\lambda\kappa}v_\mu{\rm tr}[\bar{S}_\nu F_{\lambda\kappa} B_{\bar{3}}]+H.c.,
\end{align}
where $S^{\mu}_{ab}$ is composed of Dirac spinor operators,
\begin{align}
    S^{ab}_{\mu}&=-\sqrt{\frac{1}{3}}(\gamma_{\mu}+v_{\mu})
    \gamma^{5}B^{ab}+B^{*ab}_{\mu}\equiv{ B}^{ab}_{0\mu}+B^{ab}_{1\mu},\nonumber\\
    \bar{S}^{ab}_{\mu}&=\sqrt{\frac{1}{3}}\bar{B}^{ab}
    \gamma^{5}(\gamma_{\mu}+v_{\mu})+\bar{B}^{*ab}_{\mu}\equiv{\bar{B}}^{ab}_{0\mu}+\bar{B}^{ab}_{1\mu},
\end{align}
and the  the bottomed baryon matrices are defined as
\begin{align}
B_{\bar{3}}&=\left(\begin{array}{ccc}
0&\Lambda^+_c&\Xi_c^+\\
-\Lambda_c^+&0&\Xi_c^0\\
-\Xi^+_c&-\Xi_c^0&0
\end{array}\right),\quad
B=\left(\begin{array}{ccc}
\Sigma_c^{++}&\frac{1}{\sqrt{2}}\Sigma^+_c&\frac{1}{\sqrt{2}}\Xi'^+_c\\
\frac{1}{\sqrt{2}}\Sigma^+_c&\Sigma_c^0&\frac{1}{\sqrt{2}}\Xi'^0_c\\
\frac{1}{\sqrt{2}}\Xi'^+_c&\frac{1}{\sqrt{2}}\Xi'^0_c&\Omega^0_c
\end{array}\right). \nonumber\\
B^*&=\left(\begin{array}{ccc}
\Sigma_c^{*++}&\frac{1}{\sqrt{2}}\Sigma^{*+}_c&\frac{1}{\sqrt{2}}\Xi^{*+}_c\\
\frac{1}{\sqrt{2}}\Sigma^{*+}_c&\Sigma_c^{*0}&\frac{1}{\sqrt{2}}\Xi^{*0}_c\\
\frac{1}{\sqrt{2}}\Xi^{*+}_c&\frac{1}{\sqrt{2}}\Xi^{*0}_c&\Omega^{*0}_c
\end{array}\right).\label{MBB}
\end{align}

The explicit forms of the Lagrangians can be written as,
\begin{align}
{\cal L}_{BB\mathbb{P}}&=-i\frac{3g_1}{4f_\pi\sqrt{m_{\bar{B}}m_{B}}}~\epsilon^{\mu\nu\lambda\kappa}\partial^\nu \mathbb{P}~
\sum_{i=0,1}\bar{B}_{i\mu} \overleftrightarrow{\partial}_\kappa B_{j\lambda},\nonumber\\
{\cal L}_{BB\mathbb{V}}&=-\frac{\beta_S g_V}{2\sqrt{2m_{\bar{B}}m_{B}}}\mathbb{V}^\nu
 \sum_{i=0,1}\bar{B}_i^\mu \overleftrightarrow{\partial}_\nu B_{j\mu}\nonumber\\
&-\frac{\lambda_S
g_V}{\sqrt{2}}(\partial_\mu \mathbb{V}_\nu-\partial_\nu \mathbb{V}_\mu) \sum_{i=0,1}\bar{B}_i^\mu B_j^\nu,\nonumber\\
{\cal L}_{BB\sigma}&=\ell_S\sigma\sum_{i=0,1}\bar{B}_i^\mu B_{j\mu},\nonumber\\
    {\cal L}_{B_{\bar{3}}B_{\bar{3}}\mathbb{V}}&=-\frac{g_V\beta_B}{2\sqrt{2m_{\bar{B}_{\bar{3}}}m_{B_{\bar{3}}}} }\mathbb{V}^\mu\bar{B}_{\bar{3}}\overleftrightarrow{\partial}_\mu B_{\bar{3}},\nonumber\\
{\cal L}_{B_{\bar{3}}B_{\bar{3}}\sigma}&=i\ell_B \sigma \bar{B}_{\bar{3}}B_{\bar{3}},\nonumber\\
{\cal L}_{BB_{\bar{3}}\mathbb{P}}
    &=-i\frac{g_4}{f_\pi} \sum_i\bar{B}_i^\mu \partial_\mu \mathbb{P} B_{\bar{3}}+{\rm H.c.},\nonumber\\
{\cal L}_{BB_{\bar{3}}\mathbb{V}}    &=\frac{g_\mathbb{V}\lambda_I}{\sqrt{2m_{\bar{B}}m_{B_{\bar{3}}}}} \epsilon^{\mu\nu\lambda\kappa} \partial_\lambda \mathbb{V}_\kappa\sum_i\bar{B}_{i\nu} \overleftrightarrow{\partial}_\mu
   B_{\bar{3}}+{\rm H.c.}.
   \label{LB}
\end{align}
The masses of  particles involved in the calculation are chosen as suggested central values in the Review of  Particle Physics  (PDG)~\cite{Tanabashi:2018oca}. The mass of broad $\sigma$ meson is chosen as 500 MeV.  The  coupling constants involved are listed in Table~\ref{coupling}.
\renewcommand\tabcolsep{0.13cm}
\renewcommand{\arraystretch}{1.2}
\begin{table}[h!]
\caption{The coupling constants adopted in the
calculation, which are cited from the literature~\cite{Chen:2019asm,Liu:2011xc,Isola:2003fh,Falk:1992cx}. The $\lambda$ and $\lambda_{S,I}$ are in the units of GeV$^{-1}$. Others are in the units of $1$.
\label{coupling}}
\begin{tabular}{cccccccccccccccccc}\bottomrule[2pt]
$\beta$&$g$&$g_V$&$\lambda$ &$g_{s}$\\
0.9&0.59&5.9&0.56 &0.76\\\hline
$\beta_S$&$\ell_S$&$g_1$&$\lambda_S$ &$\beta_B$&$\ell_B$ &$g_4$&$\lambda_I$\\
-1.74&6.2&-0.94&-3.31&$-\beta_S/2$&$-\ell_S/2$&$3g_1/{(2\sqrt{2})}$&$-\lambda_S/\sqrt{8}$ \\
\bottomrule[2pt]
\end{tabular}
\end{table}

With the vertices obtained from   above Lagrangians, the potential of couple-channel   interaction can be constructed easily with the help of the standard Feynman rules.
Because six channels are involved in the current work, it is
tedious and fallible  to give explicit 36 potential elements and input them into code.  Instead, in this work, following the method in Refs.~\cite{He:2019rva}, we input vertices $\Gamma$ and  propagators $P$  into  the code directly.  The potential can be written as
\begin{equation}%
{\cal V}_{\mathbb{P},\sigma}=f_I\Gamma_1\Gamma_2 P_{\mathbb{P},\sigma}f(q^2),\ \
{\cal V}_{\mathbb{V}}=f_I\Gamma_{1\mu}\Gamma_{2\nu}  P^{\mu\nu}_{\mathbb{V}}f(q^2),\label{V}
\end{equation}
The propagators are defined as usual as
\begin{equation}%
P_{\mathbb{P},\sigma}= \frac{i}{q^2-m_{\mathbb{P},\sigma}^2},\ \
P^{\mu\nu}_\mathbb{V}=i\frac{-g^{\mu\nu}+q^\mu q^\nu/m^2_{\mathbb{V}}}{q^2-m_\mathbb{V}^2},
\end{equation}
where the form factor $f(q^2)$ is adopted to compensate the off-shell effect of exchanged meson as $f(q^2)=e^{-(m_e^2-q^2)^2/\Lambda_e^2}$
with $m_e$ being the $m_{\mathbb{P},\mathbb{V},\sigma}$ and $q$ being the momentum of the exchanged  meson.
The $f_I$ is the flavor factor for certain meson exchange of certain interaction, and the explicit values are listed in Table~\ref{flavor factor}.
\renewcommand\tabcolsep{0.18cm}
\renewcommand{\arraystretch}{1.2}
\begin{table}[h!]
\caption{The flavor factors $f_I$ for certain meson exchanges of certain interaction. The values in bracket are for the case of $I=1$ if the values are different from these of $I=0$  \label{flavor factor}}
\begin{tabular}{c|ccccc}\bottomrule[2pt]
& $\pi$&$\eta$&$\rho$ & $\omega$ & $\sigma$ \\\hline
$\bar{D}^{(*)}\Xi^{(',*)}_c \to \bar{D}^{(*)}\Xi^{(',*)}_c$  &$-\frac{3}{4}[\frac{1}{4}]$ &$-\frac{1}{12}$ &$-\frac{3}{4}[\frac{1}{4}]$&$\frac{1}{4}$ & 1\\
$\bar{D}^{(*)}\Xi_c\to \bar{D}^{(*)}\Xi_c$  &$0$ &$0$ &$-\frac{3}{2}[\frac{1}{2}]$&$\frac{1}{2}$ & 2\\
$\bar{D}^{(*)}\Xi_c\to \bar{D}^{(*)}\Xi_c^{(',*)}$& $-\frac{3}{2\sqrt{2}}[\frac{1}{2\sqrt{2}}]$&$-1\over{2\sqrt{2}}$&$-\frac{3}{2\sqrt{2}}[\frac{1}{2\sqrt{2}}]$&$1\over{2\sqrt{2}}$ & 0\\
\toprule[2pt]
\end{tabular}
\end{table}

With the potential kernel obtained, we adopt the qBSE to solve the scattering amplitude~\cite{He:2014nya,He:2015mja,He:2012zd,He:2015yva,He:2017aps}. After partial-wave decomposition and spectator quasipotential approximation,  the 4-dimensional Bethe-Salpeter equation  in the Minkowski space can be reduced to a 1-dimensional  equation with fixed spin-parity $J^P$ as~\cite{He:2015mja},
\begin{align}
i{\cal M}^{J^P}_{\lambda'\lambda}({\rm p}',{\rm p})
&=i{\cal V}^{J^P}_{\lambda',\lambda}({\rm p}',{\rm
p})+\sum_{\lambda''}\int\frac{{\rm
p}''^2d{\rm p}''}{(2\pi)^3}\nonumber\\
&\cdot
i{\cal V}^{J^P}_{\lambda'\lambda''}({\rm p}',{\rm p}'')
G_0({\rm p}'')i{\cal M}^{J^P}_{\lambda''\lambda}({\rm p}'',{\rm
p}),\quad\quad \label{Eq: BS_PWA}
\end{align}
where the sum extends only over nonnegative helicity $\lambda''$.
The $G_0({\rm p}'')$ is reduced from the 4-dimensional  propagator under quasipotential approximation as  $G_0({\rm p}'')=\delta^+(p''^{~2}_h-m_h^{2})/(p''^{~2}_l-m_l^{2})$ with $p''_{h,l}$ and $m_{h,l}$ being the momenta and masses of heavy or light constituent particles.
The partial wave potential is defined with the potential of  interaction obtained in the above in Eq.~(\ref{V}) as
\begin{align}
{\cal V}_{\lambda'\lambda}^{J^P}({\rm p}',{\rm p})
&=2\pi\int d\cos\theta
~[d^{J}_{\lambda\lambda'}(\theta)
{\cal V}_{\lambda'\lambda}({\bm p}',{\bm p})\nonumber\\
&+\eta d^{J}_{-\lambda\lambda'}(\theta)
{\cal V}_{\lambda'-\lambda}({\bm p}',{\bm p})],
\end{align}
where $\eta=PP_1P_2(-1)^{J-J_1-J_2}$ with $P$ and $J$ being parity and spin for system, $D^{(*)}$ meson or $\Xi_c^{(*,')}$ baryon. The initial and final relative momenta are chosen as ${\bm p}=(0,0,{\rm p})$  and ${\bm p}'=({\rm p}'\sin\theta,0,{\rm p}'\cos\theta)$. The $d^J_{\lambda\lambda'}(\theta)$ is the Wigner d-matrix.
We also adopt an  exponential
regularization  by introducing a form factor into the propagator as~\cite{He:2015mja}
$G_0({\rm p}'')\to G_0({\rm p}'')\left[e^{-(p''^2_l-m_l^2)^2/\Lambda_r^4}\right]^2$ with $\Lambda_r$ being a cutoff.

\section{Results and discussions}\label{3}
With the  preparation above, we can perform numerical calculation to study the molecular states from the $\Xi'_c \bar{D}^{(*)}$ and $\Xi^{(*)}_c \bar{D} ^{(*)}$ interactions.After transformation of the qBSE into a matrix equation, the scattering amplitude can be obtained, and the molecular states can be searched for as the poles in the complex energy plane. As said in the Introduction, the parameters of the Lagrangians in the current work are chosen as  same as those in our previous study of the hidden-charm pentaquarks~\cite{He:2019ify,He:2019rva}. The only parameters are cutoff $\Lambda_e$ and $\Lambda_r$, which are rewritten as a form of $\Lambda_e=\Lambda_r=m+\alpha~0.22$ GeV. Hence, in the current work, only one parameter is involved, and its value will be discussed later. In the followings,  the single-channel results will be presented first. Then, the coupled-channel effects are included to produce the width of the molecular states. The couplings of molecular states to the considered channels  are also discussed to analyze the contributions of decay channels.

\subsection{Single-channel results }

\begin{figure}[htpb!]
  \centering
 \includegraphics[scale=0.65,bb=10 40 330 410,clip]{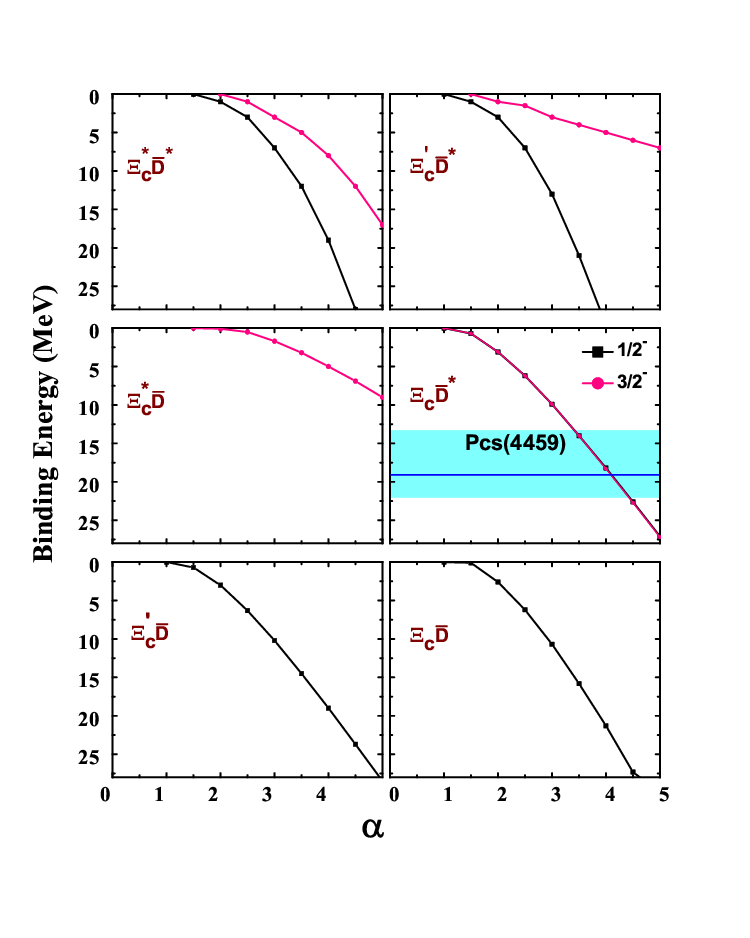}\\
  \caption{The  binding energies of the bound states from six single-channel interactions with the variation of parameter $\alpha$. All states carry scalar isospin $I=0$.  The thresholds of the six channels are 4654.6, 4587.4, 4513.2, 4478.0, 4446.0, 4336.6~MeV for $\Xi_c^*\bar{D}^*$, $\Xi'_c\bar{D}^*$, $\Xi^*_c\bar{D}$, $\Xi_c\bar{D}^*$, $\Xi'_c\bar{D}$, and $\Xi_c\bar{D}$, respectively. The blue line and the band are the experimental mass and uncertainties of the $P_{cs}(4459)$. }\label{0}
\end{figure}

 For the six channels considered, the results of single-channel calculation  are shown in  Fig.~\ref{0}. All states which can be produced in S wave with isospin $I=0$ and 1 are considered in our calculation, that is, 20 possible molecular states will be considered.

As shown in the figure, five isoscalar bound states with spin parity $1/2^-$ are produced from the $\Xi^*_c\bar{D}^*$, $\Xi'_c\bar{D}^*$, $\Xi_c\bar{D}^*$, $\Xi'_c\bar{D}$, and $\Xi_c\bar{D}$ interactions, and four isoscalar bound states with spin parity  $3/2^-$ from $\Xi^*_c\bar{D}^*$, $\Xi'_c\bar{D}^*$, $\Xi^*_c\bar{D}$, and $\Xi_c\bar{D}$ interactions. In other words, except the $\Xi_c^*\bar{D}^*$ interaction with $5/2$, bound states are produced from all other S-wave isoscalar interactions.  The states appears at $\alpha$ values of about 1, and the bound energies become deeper with increasing of the $\alpha$.  All bound states produced from the six channels considered are isoscalar. The isovector bound states with $I=1$ are also searched but found difficult to be produced, even with an $\alpha$ value of about 9.  It can be understood through the flavor factors in Table~\ref{flavor factor}. The signs of the flavor factors  for isovector states reverse compared with these for isoscalar states.  It is also found in the studies about the hidden-charm and hidden-bottom pentaquarks~\cite{Zhu:2020vto}.

 The LHCb experiment suggested that the $P_{cs}(4459)$ have a mass of about 19~MeV lower than the threshold of $\Xi_c\bar{D}^{*}$, which is also illustrated in Fig.~\ref{0} with the experimental uncertainty. To reproduce the experimental mass, the the value of $\alpha$ about 3 should be chosen.  Unfortunately, the two $\Xi_c\bar{D}^{*}$  states with $J^P=\frac{1}{2}^-$ and $\frac{3}{2}^-$  have  very close binding energies with single-channel calculation.  Hence, we can not determine the spin parities the $P_{cs}(4459)$  in the single-channel calculation. The explicit analysis suggests that the contribution of vector exchanges are not sensitive to the spin, and the pseudoscalar exchanges are absent in this channel as shown in Table~\ref{flavor factor}. It can be expected that after the pseudoscalar exchanges involved in coupled-channel effect, these two states will split.

\subsection{Coupled-channel results and position of poles}\label{4}

The single-channel calculation can provide a basic picture of the bound states from six interactions considered.  Due to lack of couplings between different channels, all states are bound states, that is, the poles are at the real axis.  In the followings, we will include the coupled-channel effect into the single-channel calculation. The coupled-channel results for the molecular states are  listed in Table~\ref{six}. Here we label a pole by nearest threshold and its spin.
\renewcommand\tabcolsep{0.25cm}
\renewcommand{\arraystretch}{1.1}
\begin{table}[h!]
\caption{The molecular states with coupled-channel calculation. Positions are given by the corresponding threshold subtracted by the position of a pole, $M_{th}-z$,  in the unit of MeV. The $\alpha$ and the pole are in the units of GeV, and MeV, respectively. } \label{six}
{	\begin{tabular}{c|rr|rr}\bottomrule[2pt]

$\alpha$& $\Xi_c^{*}\bar{D}^{*}\frac{1}{2}$&$\Xi_c^{*}\bar{D}^{*}\frac{3}{2}$&$\Xi'_c\bar{D}^{*}\frac{1}{2}$ & $\Xi'_c\bar{D}^{*}\frac{3}{2}$  \\\hline
$2.0$  &$1.5+0.4i$& $0.2+0.4i$&$5.2+1.9i$&$1.3+2.1i$ \\

$2.5$  &$3.7+1.4i$& $2.0+4.4i$&$6.9+4.6i$&$1.9+5.0i$\\

$3.0$ &$7.4+3.3i$& $      N                 $&$8.2+6.8i$&$2.6+6.9i$ \\

$3.5$ &$11.4+6.6i$& $       N              $&$4.9+8.7i$&$5.6+7.3i$ \\\hline

%$4.0$ &$19.2+10.7i $& $                     $&$7.4+8.0i$&$10.5+11.1i$ \\

%$4.5$ &$29.3+15.3i $& $                     $&$13.3+9.4i$&$16.2+15.0i$ \\

$\alpha$& $--$& $\Xi_c^{*}\bar{D}\frac{3}{2}$& $\Xi_c\bar{D}^{*}\frac{1}{2}$ &$\Xi_c\bar{D}^{*}\frac{3}{2}$\\\hline
$2.0$ &$--$&$0.1+2.1i$&$2.2+0.9i$& $4.6+0.7i$\\
 $2.5$ &$--$&$2.8+4.3i$&$4.5+2.2i$& $10.5+1.1i$\\
$3.0$ &$--$&$6.3+7.0i$&$\textbf{7.9+4.0i}$& $\textbf{19.7+1.6i}$\\
$3.5$&$--$&$18.4+9.3i$&$13.3+6.4i$& $33.3+0.7i$\\\hline
%$4.0$&$--$&$34.2+5.9i$&$23.6+7.5$& $51.8+2.0$\\
%$4.5$&$--$&$44.8+1.2i$&$32.0+0.0$& $79.0+3.8$\\
$\alpha$&$\Xi'_c\bar{D}\frac{1}{2}$ & $--$ & $\Xi_c\bar{D}\frac{1}{2}$& $--$  \\\hline

$2.0$  &$4.9+0.0i$&$    --      $ &$0.2$&$   --       $\\

$2.5$  &$10.8+0.0i$&$    --      $ &$3.9$&$    --      $\\

$3.0$ &$19.4+0.0i$&$       --   $ &$8.5$&$    --      $\\

$3.5$ &$ 30.9+0.0i$&$    --     $ &$14.0$&$   --       $\\

%$4.0$ &$46.1+0.0$&$     --     $ &$20.8+0.0$&$   --       $\\
%
%$4.5$ &$65.6+0.0$&$       --    $ &$29.6+0.0$&$    --      $\\
\toprule[2pt]
\end{tabular}}
\end{table}

\begin{figure}[h!]
  \centering
  \includegraphics[scale=0.76, bb=20 0 500 248,clip]{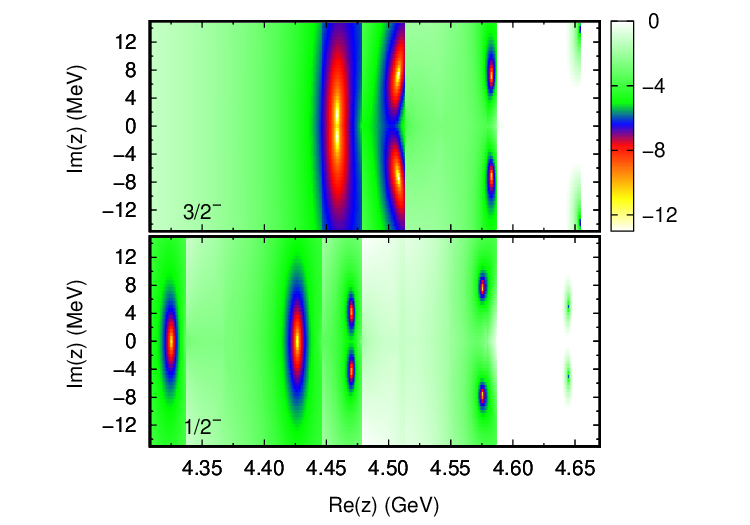}\\
  \caption{The $\log|1-V(z)G(z)|$ with the variation of  $z$ for the $\Xi'_c \bar{D}^{(*)}$ and $\Xi^{(*)}_c \bar{D}^{(*)}$ interaction with $J^P=1/2^-$ and $3/2^-$  at $\alpha=3.0$. The  color  means the value of $\log|1-V(z)G(z)|$ as shown in the color box.}\label{3.0}
\end{figure}

As listed in Table~\ref{six},  the poles leave the real axis after the coupled-channel effects are included, and acquire imaginary parts, which related to the width as $\Gamma=2 {\rm Im}z$. As in the single-channel calculation, two poles near $\Xi_c\bar{D}^*$ threshold can be produced with spin parties $1/2^-$ and $3/2^-$, respectviely. With an $\alpha$ value of about 3,  the masses of these two states are close to the experimental observed values of $P_{cs}(4459)$. The width of these two state, 8.0 and 3.2 MeV, are smaller than the experimental values, which may be due to the possible two-peak structure of the $P_{cs}(4459)$.   To give a visual picture of the results,  the explicit results with $\alpha=3$  are presented in Fig.~\ref{3.0}, where the positions of the poles are illustrated clearly in the complex energy plane.

As shown in Fig.~\ref{3.0}, among all states, the state $\Xi_c\bar{D}^{*}(\frac{3}{2}^-)$  are most obvious and has largest coverage in the complex energy plane. It should be easier to be observed in the experiment.  At the $\alpha$ of 3.0, the mass of state $\Xi_c\bar{D}^{*}(\frac{3}{2}^-)$ is $4458.3$MeV, quite closing to the experimental value of the $P_{cs}(4459)$. It indicts that the  $\Xi_c\bar{D}^{*}(\frac{3}{2}^-)$ state is the most possible candidate of the $P_{cs}(4459)$.   As expected, the inclusion of the coupled-channel effect makes the two states from the $\Xi_c\bar{D}^{*}$ interaction deviate from each other. The state with spin parity $3/2^-$ moves further  while the state with $(1/2^-)$ closer to the $\Xi_c\bar{D}^{*}$ threshold. The higher state $\Xi_c\bar{D}^{*}(1/2^-)$ has a mass of $4470.1$MeV.  Though two states are separated  by the coupled-channel effect, the small mass gap about 10 MeV  requires high precision measurement to distinguish  these two states  in experiment.

With the $P_{cs}(4459)$ reproduced in our theoretical frame, we can provide the prediction of other possible strange hidden-charm molecular states.   Two states with spin parity $1/2^-$ are produced near the $\Xi_c\bar{D}$ and $\Xi'_c\bar{D}$  thresholds, respectively. The lowest state has zero width because no channel lower than the $\Xi_c\bar{D}$ threshold is considered in the current work. The state near the $\Xi'_c\bar{D}$ threshold has a negligible width due to its very weak coupling to the $\Xi_c\bar{D}$ channel. No $3/2^-$ state is given here because we only consider the state which can be produced in S wave.
A state can be found near the $\Xi^{*}_c\bar{D}$ threshold with spin parity $3/2^-$ with a width of about 14 MeV at $\alpha$ of a value 3 as shown in Table~\ref{six}.  Two poles can also be  found near the $\Xi'_c\bar{D}^{*}$ threshold with spin parities $1/2^-$ and $3/2^-$. The widths of this two states are similar, and the mass gap is about 5 MeV.  Near the highest threshold $\Xi_c^*\bar{D}^*$, two states can be also  produced from the interaction. However, as two $\Sigma_c^*\bar{D}^*$ states predicted in the hidden-charm sector~\cite{He:2019rva}, the signal of these two states are very weak as shown in Fig.~\ref{3.0}.  As shown in Table~\ref{six}, with a larger cutoff, the pole of state with $3/2^-$ is even difficult to be read out from the results. These two states may be  difficult to be observed in experiment.

\subsection{The widths of the molecular states from each decay channel}

In the above, we present the  widths of the states with all channels considered.
The width of a molecular state is from decaying into the channels with thresholds lower than the mass of the state. The width from one of decay channels of a molecular state shows the strength of the couplings between the molecular state and the corresponding channel.  By comparing the single and coupled channel results, one can find the coupled-channel effects are not very large. Hence, the main production channel of a molecular can be easily determined by the nearest threshold.  In Table~\ref{pole},  we present the pole obtained by a two-channel calculation with main production channel and a decay channel to show the coupling of a molecular state to the decay channel.
\renewcommand\tabcolsep{0.1cm}
\renewcommand{\arraystretch}{1.1}
\begin{table}[h!]   %hpbt!
\caption{The positions of poles of the molecular states with two-channel calculation.  Positions are given by the corresponding threshold subtracted by the position of a pole, $M_{th}-z$,  in the unit of MeV. The $\alpha$ is in the unit of GeV.
}\label{pole}
\begin{tabular}{c|rrrrr}\toprule[2pt]
 $\alpha$ & \multicolumn{1}{c}{$\Xi'_c \bar{D}^*$} & \multicolumn{1}{c}{$\Xi^*_c \bar{D}$} &  \multicolumn{1}{c}{ $\Xi_c \bar{D}^*$}&   \multicolumn{1}{c}{$\Xi'_c \bar{D}$} &  \multicolumn{1}{c}{  $\Xi_c \bar{D}$} \\
\hline
%1
\multicolumn{6}{c}{$\Xi^*_c \bar{D}^*(1/2^-) \quad  M_{th}=4654.6$MeV}\\
\hline
$2.0$       &$1.7+0.1i$    &$ 1.8+0.1i$          &$1.8+0.0i$          &$1.7+0.2i$     &$1.8+0.0i$  \\
$2.5$      &$4.8+0.2i$    &$ 4.9+0.2i$         &$4.6+0.1i$         &$4.2+0.5i$      &$4.5+0.0i$    \\
$3.0$       &$9.0+0.4i$    &$ 9.2+0.5i$         &$8.3+0.2i$        &$7.4+1.4i$      &$7.6+0.1i$    \\
$3.5$   &$14.0+0.6i$  &$14.5+0.8 i$     &$12.9+0.3i$       &$11.3+2.8i$    &$11.3+0.2i$   \\
\hline
%2
\multicolumn{6}{c}{$\Xi^*_c \bar{D}^*(3/2^-) \quad  M_{th}=4654.6$MeV}\\
\hline
$2.0$       &$0.2+0.1i$    &$ 0.4+0.0i$          &$0.1+0.2i$          &$0.4+0.0i$     &$0.4+0.0i$    \\
$2.5$      &$1.2+0.7i$    &$ 1.5+0.3i$         &$0.1+0.3i$         &$2.0+0.1i$      &$1.9+0.0i$     \\
$3.0$                   &$2.7+1.4i$    &$ 2.7+0.8i$         &$0.1+2.5i$         &$4.2+0.1i$      &$4.1+0.0i$     \\
$3.5$                    &$4.3+2.3i$  &$3.7+1.7 i$          &$0.1+3.1i$          &$6.9+0.1i$     &$6.7+0.0i$   \\
\hline
%3
\multicolumn{6}{c}{$\Xi'_c \bar{D}^*(1/2^-)\quad  M_{th}=4587.4$MeV}\\
\hline
$2.0$        &$--$    &$ 5.8+0.1i$          &$4.7+1.5i$          &$5.7+0.1i$     &$5.7+0.0i$    \\
$2.5$     &$--$    &$ 10.9+0.4i$         &$8.1+3.8i$         &$10.4+0.2i$      &$10.1+0.2i$   \\
$3.0$       &$--$    &$ 17.3+0.9i$         &$12.1+6.9i$        &$15.6+0.6i$      &$14.9+0.7i$    \\
$3.5$       &$--$    &$24.7+2.0 i$     &$16.3+12.2i$       &$20.8+1.3i$    &$19.8+1.6i$  \\
\hline
%4
\multicolumn{6}{c}{$\Xi'_c \bar{D}^*(3/2^-)\quad  M_{th}=4587.4$MeV}\\
\hline
$2.0$       & $--$      &$ 1.6+0.0i$          &$0.7+2.1i$          &$1.8+0.0i$     &$1.3+0.3i$   \\
$2.5$       &$--$      &$ 3.2+0.1i$         &$1.3+5.9i$            &$3.6+0.0i$      &$1.8+1.4i$     \\
$3.0$      &$--$       &$ 4.9+0.4i$         &$--$                          &$5.4+0.1i$      &$--$    \\
$3.5$      &$--$       &$6.2+0.6i$            &$--$                       &$7.1+0.1i$       &$--$   \\
\hline
%5
\multicolumn{6}{c}{$\Xi^*_c \bar{D}(3/2^-)\quad  M_{th}=4513.2$MeV}\\
\hline
$2.0$       &$--$    &$ --$          &$0.1+1.9i$          &$0.1+0.0i$     &$0.1+0.0i$    \\
$2.5$      &$--$    &$ --$         &$0.7+3.7i$         &$0.5+0.0i$      &$0.5+0.0i$    \\
$3.0$      &$--$    &$ --$         &$3.9+5.2i$        &$1.7+0.0i$      &$1.7+0.0i$     \\
$3.5$    &$--$      &$--$         &$11.2+6.1i$       &$3.3+0.0i$    &$3.2+0.0i$   \\
\hline
%6
\multicolumn{6}{c}{$\Xi_c \bar{D}^*(1/2^-)\quad  M_{th}=4478.0$MeV}\\
\hline
$2.0$      &$--$    &$ --$          &$--$          &$2.1+0.7i$     &$3.1+0.0i$    \\
$2.5$      &$--$    &$ --$         &$--$         &$4.1+1.6i$      &$6.2+0.0i$     \\
$3.0$       &$--$    &$ --$         &$--$        &$6.3+2.7i$      &$9.9+0.0i$     \\
$3.5$    &$--$    &$--$         &$--$       &$8.6+3.8i$    &$14.0+0.0i$   \\
\hline
%7
\multicolumn{6}{c}{$\Xi_c \bar{D}^*(3/2^-)\quad  M_{th}=4478.0$MeV}\\
\hline
$2.0$     &$--$    &$ --$          &$--$          &$3.9+0.6i$     &$3.1+0.0i$  \\
$2.5$       &$--$    &$ --$         &$--$         &$8.2+0.9i$      &$6.2+0.0i$    \\
$3.0$        &$--$    &$ --$         &$--$        &$13.7+1.1i$      &$9.9+0.0i$     \\
$3.5$   &$--$    &$--$         &$--$       &$20.5+1.2i$    &$14.0+0.0i$   \\
\hline
%8
\multicolumn{6}{c}{$\Xi'_c \bar{D}(1/2^-)\quad  M_{th}=4446.0$MeV}\\
\hline
$2.0$      &$--$    &$--$          &$--$          &$--$     &$3.0+0.0i$   \\
$2.5$        &$--$    &$--$          &$--$          &$--$     &$6.3+0.0i$    \\
$3.0$       &$--$    &$--$          &$--$          &$--$     &$10.2+0.0i$   \\
$3.5$     &$--$    &$--$          &$--$          &$--$     &$14.5+0.0i$   \\
\bottomrule[2pt]
\end{tabular}
\end{table}

In the table, the results for the $\Xi_c^*\bar{D}^*$ state near the highest threshold  with $1/2^-$ are listed first. Among five decay channels listed in the second to sixth columns,  the  strongest coupling is found in $\Xi'\bar{D}$ channel. The $\Xi_c\bar{D}^*$ is the most important decay channel of the $\Xi_c^*\bar{D}^*$ state with $3/2^-$, and the $\Xi'_c\bar{D}^*$ and $\Xi^*\bar{D}$ channels also have considerable contribution to its width.  The $\Xi_c \bar{D}^* $ channel is also the main decay channel of the molecular states  $\Xi^{'}_c \bar{D}^{*} ({3}/{2}^-, {3}/{2}^-)$ and $\Xi^{*}_c \bar{D}( {3}/{2}^{-}) $. For the $\Xi_c\bar{D}^*$ states with $1/2^-$ and $3/2^-$, the $\Xi'\bar{D}$ channel is dominant to produce their widths.  The $\Xi'_c \bar{D}(1/2^-)$ state couples to its only open channel $\Xi_c\bar{D}$ very weakly.  Hence, the  $\Xi_c\bar{D}^*$ and $\Xi'\bar{D}$ channel are important for the decays of four and three molecular states, respectively.

\section{Summary}\label{5}

In this work, the strange hidden-charm pentaquarks are studied in the molecular picture in the qBSE approach. The newly observed state $P_{cs}(4459)$ could be interpreted as the $\Xi_c \bar{D}^{*} (\frac{3}{2}^-)$ molecular state, which has strong coupling to $\Xi'_c\bar{D}$ channel.  The pole of this state is much more obvious in the complex energy plane than other states, which may be the reason that it was observed first. Our result does not exclude the possibility that the  $P_{cs}(4459)$ is  a two-peak structure composed of the $\Xi_c \bar{D}^{*}$ states with $1/2^-$ and $3/2^-$, which is helpful to understand the experimental width. The current work is performed in the same theoretical frame as previous study of the hidden-charm pentaquarks with almost the same parameters~\cite{He:2019ify}.  Hence,  the four hidden-charm pentaquarks, $P_c(4312)$, $P_c(4380)$, $P_c(4440)$, $P_c(4457)$, and the strange hidden-charm pentaquark $P_{cs}(4459)$ can be interpreted well in the molecular picture and assigned as S-wave state from the interaction of the (strange) charmed baryon and anti-charmed meson.

In the same model, we also predict other possible strange hidden-charm pentaquarks.  Two states with spin parity $1/2^-$ are predicted near the $\Xi'_c\bar{D}$ and $\Xi_c\bar{D}$ thresholds, respectively. Since the decay channels of these two states are less than other states, their widths may be smaller.  Near the $\Xi^*_c\bar{D}$ threshold, an obvious pole can be found in the complex energy plane with $3/2^-$. This state couples strongly with $\Xi_c\bar{D}$ channel.  As two states near the $\Xi_c \bar{D}^{*}$ threshold, one can find two states with $1/2^-$ and $3/2^-$  near the $\Xi'_c\bar{D}^*$ threshold, which have strong coupling to the $\Xi_c \bar{D}^{*}$ channel.  The  poles can also be found near the highest threshold $\Xi_c^*\bar{D}^*$. Such poles are dimly in the complex energy plane, and may be difficult to be find in experiment.  The experimental  research of those states are helpful to understand the origin of the $P_{cs}$ and $P_c$ states.

\vskip 10pt

\noindent {\bf Acknowledgement} This project is supported by the National Natural Science
Foundation of China (Grants No. 11675228).

%

%\bibliography{../../../reference/Jabref/bibliography}

\end{document}